\documentclass[aps,prb,twocolumn]{revtex4-1}
\usepackage{graphicx}
\usepackage{bm}
\usepackage{amsmath}
\begin{document} 
\title{Bond particle theory for the pseudogap phase of underdoped cuprates}
\author{R. Eder}
\affiliation{Karlsruher Institut f\"ur Technologie,
Institut f\"ur Festk\"orperphysik, 76021 Karlsruhe, Germany}
\date{\today}
\begin{abstract}
We present a theory for the lightly doped t-J model which is of possible
relevance for the normal state of underdoped cuprates.
Starting from an arbitrary dimer covering of the plane an exact representation 
of the t-J Hamiltonian in terms of bond bosons and fermions can be derived.
Since all dimer coverings must give identical results for observable 
quantities we construct an approximate but translationally invariant
Hamiltonian by averaging the bond particle Hamiltonian over all possible dimer 
coverings. Treating the resulting Hamiltonian in mean-field approximation
we find fermi pockets centered near $(\pi/2,\pi/2)$ with a total area of $x/2$ 
(with $x$ the hole concentration) and a gapped spin-wave-like
band of triplet excitations.
\end{abstract} 
\maketitle
\section{Introduction}
Copper oxide based superconductors have a phase diagram which is
similar to a large number of heavy fermion compounds and the
iron pnictide superconductors\cite{Keimer}. In all these materials a phase 
transition occurs at zero temperature as a function of some control parameter
which is surrounded by a superconducting dome. In the cuprates the
control parameter is the hole concentration $x$ in the CuO$_2$ planes
but whereas the phase on the overdoped side of the transition appears to
be a normal metal albeit with correlation enhanced effective mass,
the underdoped phase - the so-called pseudogap phase - is not well 
understood. To begin with, we briefly summarize some experimental results
for this phase. \\
Below the pseudogap temperature $T^*$ which decreases monotonously 
with $x$, angle resolved photoelectron spectroscopy (ARPES) shows fermi arcs 
and a pseudogap\cite{Loeser,Ding,Damascelli} i.e. unlike expected for 
the hole-doped density functional band structure, the quasiparticle band does 
not reach the chemical potential $\mu$ in a sizeable range in 
${\bm k}$-space around $(\pi,0)$. One possible explanation would be that the 
fermi arc really is one half of an elliptical or semi-elliptical hole pocket 
which is centered near $(\frac{\pi}{2}, \frac{\pi}{2})$, formed by 
a band with weak dispersion along $(0,\pi)\rightarrow (\pi,0)$. To reconcile 
this interpretation with experiment one has to make the additional assumption
that the ARPES weight of the quasiparticle band has a strong
${\bm k}$-dependence and drops sharply to almost zero upon crossing a line in 
${\bm k}$-space which roughly corresponds to the antiferromagnetic zone 
boundary, so that only the part of the pocket facing $(0,0)$ can be resolved, 
whereas the part facing $(\pi,\pi)$ has too low spectral weight. In fact, much 
the same behaviour is indeed observed in insulating 
cuprates\cite{Wells,Ronning}  
where this phenomenon has been termed the remnant fermi surface.
In the underdoped compounds the drop of spectral weight would have to be even 
more pronounced, however, to be compatible with experiment.\\
The length of the fermi arcs is independent of temperature\cite{Kondo,Yang} 
up to $T^*$, as expected for a true fermi surface. At $T^*$
the arcs disappear abruptly and ARPES apparently shows the free-electron 
fermi surface. The length of the arcs increases with $x$\cite{Tanaka,Yang}, 
which suggests that the carriers which form the pockets are the doped 
holes. A somewhat unusual feature is the temperature dependendence of the 
dispersion, i.e. while the arc length
is temperature independent, the dispersion along $(0,\pi)\rightarrow (\pi,0)$
flattens with increasing temperature\cite{Hashimoto,Yang}, so that the 
pseudogap seems to close with increasing temperature. It is plausible
that the motion of the doped holes through the `spin background' will be 
influenced by the spin correlations of the latter. Since these change with 
temperature, a temperature dependence of the hole dispersion is not entirely 
unexpected.\\
The asymmetry of the spectral weight around 
$(\frac{\pi}{2}, \frac{\pi}{2})$ which gives rise to the remnant fermi surface 
in the insulating compounds is reproduced by exact diagonalization of the 
half-filled t-J or Hubbard model\cite{Eskes}. It can be explained by the 
coupling of the photohole to the quantum spin fluctuations of the Heisenberg 
antiferromagnet or t-J model\cite{specpaper}, further enhanced by the coupling 
to charge fluctuations in the Hubbard model\cite{Oleg}. \\
Further insight can be gained from thermodynamical and transport properties.
Thereby an extra complication has to be taken into account, namely the
tendency of underdoped cuprates to form inhomogeneous states with
charge density wave (CDW) order with ordering wave vector ${\bf q}_{CDW}=(q,0)$,
the precise nature of which depends on material. It may be combined with spin 
order\cite{Tranquada}, `checkerboard order'\cite{Howald}, or CDW order without 
spin order\cite{Ghiringhelli}.
Typically these ordered states are observed at low temperatures and in a 
certain doping range  $[x_{min}^{CDW}, x_{max}^{CDW}]$.
When this is taken into account various results indicate
that the underdoped cuprates are fermi liquids. The charge carrier 
relaxation rate  $\tau^{-1}$ extracted from the optical conductivity of various
underdoped cuprates (with $x\approx 0.1$) has a quadratic dependence on both, 
frequency $\omega$ and temperature $T$\cite{optical}. 
For compounds with $x<0.15$ and for temperatures below $T^*$,
the dc-resistivity $\rho$ varies with temperature as $\rho=A\cdot T^2$ 
with an $x$-dependent constant $A$\cite{Ando,sheet}, although this behaviour 
is masked at
lower temperatures by CDW order, superconducting fluctuations or 
localization\cite{sheet}. This is also consistent with the variation of the 
Hall angle with temperature as $\cot(\Theta_H)\propto T^2$\cite{Ando}. 
The Wiedemann-Franz law is obeyed\cite{Grissonnanche} as is Kohler's 
rule\cite{Chang}. The entropy $S$ and the magnetic suceptibility $\chi$
are related over a wide doping range by $S=aT\chi$, with $a$ the Wilson ratio 
for spin-$\frac{1}{2}$ fermions\cite{Loram}. This is expected for a free fermi 
gas because $S/T$ and $\chi$ probe the fermionic density of states (DOS) 
around $\mu$ by similar weighting functions so that the proportionality should 
hold even for a structured or temperature dependent DOS\cite{Loram}.\\
The one unusual feature, however, is the $x$-dependence of transport 
quantities. It is found that $A\propto x$ as would be expected from Drude theory
for a fermi gas with carrier density $n_c=x$, and the value of $A$ per 
CuO$_2$ plane is material-independent\cite{sheet}.
For larger $x\approx 0.2$, and when superconductivity is suppressed 
by a magnetic field, the carrier density inferred from the 
$T\rightarrow 0$ limiting values of $\rho$ shows a sharp crossover 
at a material-dependent $x^*\approx0.2$ from $n_c=1+x$ for $x>x^*$ to $n_c= x$ 
for $x<x^*$\cite{Badoux,Collignon}. 
The carrier density can also be inferred from the Hall constant $R_H$
but this is complicated by the occurence of CDW order.
In the absence of CDW order $R_H>0$\cite{Badoux,Collignon} as expected
for hole-like carriers.
From measurements of $R_H$ at sufficiently high temperatures, where there is 
neither superconductivity nor CDW order, one infers $n_c=x$
at $x\le 0.1$\cite{Ong,Takagi,Padilla}. For larger $x$
and when superconductivity is suppressed by a magnetic field, $n_c$
inferred from $R_H$ again crosses sharply from $n_c=1+x$ to $n_c=x$ at
the same $x^*$ where this occurs for $\rho$\cite{Badoux,Collignon}. 
Thereby $x^*$ coincides with the `critical doping' where $T^*(x)\rightarrow 0$
as inferred from a variety of physical properties\cite{Tallon}.
It is found that $x^*>x_{max}^{CDW}$ so that the crossover in the $x$-dependence
of $n_c$ cannot be related to CDW order\cite{Badoux,Collignon}, which 
can also be seen from the fact that for $x<x_{min}^{CDW}$ the behaviour 
$n_c=x$ is recovered\cite{Badoux,Collignon}. 
In the CDW-ordered state itself quantum oscillations\cite{Doiron,Sebastian,Chan}
and quadratic temperature dependence of $\rho$\cite{Proust}
confirm the fermi-liquid nature of the ground state whereas the negative
$R_H$ suggests the presence of electron pockets\cite{LeBoeuf}, likely caused by
a reconstruction of the hole pockets. The oberservation of spin zeros
in the angular dependence of the quantum oscillations suggests that the
carriers are spin-$\frac{1}{2}$ fermions\cite{Ramshaw}.
The wave vector ${\bf q}_{CDW}$
of the CDW modulation varies with $x$ and this variation is consistent 
with the assumption that ${\bf q}_{CDW}$ corresponds to the nesting vector 
connecting the tips of the fermi arcs\cite{Comin} along the direction 
$(\pi,0)\rightarrow (0,\pi)$.
This is to be expected in the hole pocket picture because for a band with weak 
dispersion along $(\pi,0)\rightarrow (0,\pi)$ the tips would be the part of 
the fermi surface with the lowest fermi velocity and hence the highest density
of states. Generally, assuming that the holes rather than the electrons are the
mobile fermions, one would have a system with a low density of carriers with
high effective mass and the average energy of delocalization would be
further reduced by the fact that there are four equivalent pockets so that the
fermi energy is reduced by a factor of $4$. Such a system may well be 
susceptible to charge ordering due to long-ranged Coulomb interaction.
Lastly, the Drude weight in the optical conductivity is 
$\propto x$\cite{Padilla,optical}, again consistent with $n_c=x$.
The effective mass, as inferred from $n_c$ and the optical sum-rule,
is practically independent of doping over the underdoped region,
and in fact also the antiferromagnetic phase\cite{Padilla}.\\
A simple and unifying interpretation for a large body of
ARPES and transport experiments on underdoped 
cuprates thus would be that below $T^*(x)$ these compounds are fermi 
liquids formed by spin-$\frac{1}{2}$ fermions which correspond to the 
doped holes. The fermi surface takes the form of hole 
pockets centered near $(\frac{\pi}{2},\frac{\pi}{2})$ and with a `dark side' 
towards $(\pi,\pi)$.This is also consistent with exact diagonalization studies 
for the t-J model which show that for 
$x\approx 0.1$ the fermi surface takes the form of hole pockets\cite{pockpaper} 
and that the low lying eigenstates can be mapped one-to-one to those of weakly 
interacting spin-1/2 fermions corresponding to the doped 
holes\cite{nishimoto} - which is the defining property of a fermi liquid. \\
The situation is very different for overdoped compounds where at low temperature
ARPES\cite{Feng,Chuang}, magnetoresistance\cite{Hussey},
and quantum oscillations\cite{Vignolle_over} show a 
free-electron-like fermi surface which takes the form of a 3D hole barrel
around $(\pi,\pi)$ and covers a fraction of the Brillouin zone of $(1+x)/2$. 
Consistent with the observation of quantum oscillations the transport properties
are fermi-liquid like\cite{Nakamae}. Such a fermi surface is expected in the 
limit of small electron density, $x\rightarrow 1$, so that
the range of applicability of this limit appears to extend down to $x^*$.
This is also consistent with
exact diagonalization studies of the dynamical spin and density correlation 
functions in the t-J model, which indicate a transition to a more conventional 
renormalized free-electron
fermi surface at around $x=0.25$\cite{low,intermediate}.\\
Taken together, the above experimental results suggest that the zero-temperature
phase transition in the cuprates at $x^*$ corresponds to the transition 
between the two types of fermi surfaces: from
pockets formed by the holes doped into the lower Hubbard band for $x<x^*$,  
to a more conventional fermi liquid with correlation-enhanced band mass
for $x>x^*$. While the latter phase probably may be adequately described 
by a Gutzwiller-projected fermi sea, the phase realized for $x<x^*$ is more 
elusive and it is the purpose of the present manuscript to describe a theory 
which can describe it. The goal of the present manuscript
therefore is to develop a theory for
the doped, paramagnetic Mott-insulator which is compatible with the
scenario suggested by the above experimental results: a translationally 
invariant state without any type of order but with short range 
antiferromagnetic spin correlations, which is a fermi liquid of 
spin-$\frac{1}{2}$ fermions which do correspond to the doped holes rather 
than the electrons, so that the volume of the fermi surface is proportional 
to $x$, rather than $1-x$. As will be seen below, the bond particle theory
which was used initially for the study of spin systems\cite{SachdevBhatt} 
is paricularly suited to do so, because it contains the right types of 
elementary excitations as its `natural particles'.\\
Theoretically, a hole pocket fermi surface can always be produced by backfolding
a free-electron-like fermi surface assuming some order parameter 
with wave vector $(\pi,\pi)$\cite{Chakravarty,Chakrapipi}. However, 
no evidence for such an order parameter has been observed so far. Constructing 
a theory which gives a fermi  surface with a volume proportional to $x$ without
invoking backfolding of a free-electron fermi surface is not achieved easily.
Various authors have proposed that a fluctuating rather than a static order 
order parameter is already sufficient to backfold the fermi 
surface\cite{QiSachdev,MoonSachdev,PunkSachdev,holtprl,holtprb} 
or that short-range antiferromagnetic correlations may cause the 
pseudogap\cite{Senechal,Kyung}. Hole pockets can also be produced
by a phenomenological ansatz for the self-energy\cite{RiceYangZhang} and
it has also been proposed that the origin of the pseudogap is checkerboard-type
CDW order\cite{Checkerboard}.
On the other hand one may also assume that there is no underlying
free electron fermi surface which can be backfolded. Rather, in this picture
the hole pockets are a consequence of the proximity to the Mott insulator, 
which corresponds to a nominally half-filled band but has no fermi surface at 
all.
The Hubbard-I approximation\cite{Hubbard1} predicts a hole pocket 
centered  at $(\pi,\pi)$ in a paramagnetic and translationally invariant 
ground state, whereby the volume of the pocket increases monotonically with 
$x$. However, the increase is nonlinear in $x$ which seems counterintuitive.
The Hubbard-I approximation can be reformulated as a theory for
hole-like and doublon-like quasiparticles, which results in hole pockets
with a volume that is strictly $\propto x$, and if short range antiferromagnetic
spin correlations are incorporated the pocket indeed is centered near 
$(\frac{\pi}{2},\frac{\pi}{2})$\cite{myhub}.
Another such theory is the quantum dimer model proposed by 
Punk {\em et al.}\cite{Punk,Huber,Feldmeier} and the theory 
to be outlined below has some similarity to this theory.
More precisely, in the following we apply bond particle theory to the t-J model.
This was proposed by Sachdev and Bhatt\cite{SachdevBhatt} and applied 
subsequently to spin ladders\cite{Gopalan,Sushkovladder,JureckaBrenig},
bilayers\cite{vojta1,vojta2}, intrisically dimerized spin 
systems\cite{Sushkov_doped, Park} and the `Kondo necklace'\cite{Siahatgar}.
\section{Formalism}
\subsection{Hamiltonian}
We consider the t-J model on a 2D square lattice with
$N$ sites, labeled by indices $i$, $j$ and periodic boundary conditions.
The Hamiltonian reads\cite{Chao,ZhangRice}
\begin{eqnarray*}
H = -\sum_{i,j}\;\sum_{\sigma} t_{i,j} \hat{c}_{i,\sigma}^\dagger \hat{c}_{j,\sigma}^{}
+ J\;\sum_{\langle i,j\rangle}\;{\bf S}_i\cdot {\bf S}_j,
\end{eqnarray*}
where $\hat{c}_{i,\sigma}^\dagger=c_{i,\sigma}^\dagger(1-n_{i\bar{\sigma}})$
${\bf S}_i$ is the operator of electron spin at site $i$
and $\langle i,j\rangle$ denotes a sum over all pairs of nearest
neighbors.
We assume that the hopping integrals $t_{i,j}$ are different from zero
only for nearest ($(1,0)$-like), 2$^{nd}$ nearest ($(1,1)$-like)
and 3$^{nd}$ nearest ($(2,0)$-like) neighbors, and call the respective
hopping integrals $t$, $t'$ and $t''$. The nearest neighbor $t$
will be taken as the unit of energy. 
\subsection{States of a Single Dimer}
The basic idea of the calculation is to represent eigenstates
of a single dimer by bond bosons (for even electron number)
and bond fermions (for odd electron number). 
We consider a single dimer with the sites labeled $1$ and $2$ and
first write down the four states with two electrons, 
that means the singlet and the triplet.
Introducing the matrix $4$-vector
$\bm{\gamma}=(\tau_0,\bm{\tau})$ with $\tau_0=1$ and 
$\bm{\tau}$ the vector of Pauli matrices,
the state $4$-vector $\bm{\beta}=(s, \bm{t})$ 
is\cite{SachdevBhatt,Gopalan}
\begin{eqnarray}
|\bm{\beta}\rangle =\frac{1}{\sqrt{2}}\;\sum_{\sigma,\sigma'}\;
c_{1,\sigma}^\dagger \;(\bm{\gamma}i\tau_y)_{\sigma,\sigma'}\;c_{2,\sigma'}^\dagger
|0\rangle.
\label{twoestates}
\end{eqnarray}
The four states with a single electron can be classified by their parity
and $z$-spin\cite{Sushkovladder,Sushkov_doped}
\begin{eqnarray}
 |f_{\pm,\sigma}\rangle &=&
\frac{1}{\sqrt{2}}\;(c_{1,\sigma}^\dagger \pm c_{2,\sigma}^\dagger)|0\rangle.
\label{oneestates}
\end{eqnarray}
The last eigenstate is the empty dimer $|e\rangle=|0\rangle$.
If the dimer is in one of the states with two electrons,
(\ref{twoestates}), we consider it as
occupied by a boson, created by the operator $4$-vector
$(s^\dagger,{\bf t}^\dagger)$, whereas a dimer in one of the states
(\ref{oneestates}) is considered as occupied by a fermion, created by
$f_{\pm,\sigma}^\dagger$. If the dimer is empty we again
consider it as occupied by a boson, created by $e^\dagger$. 
The states $|s\rangle$, $|f_{+,\sigma}\rangle$ and $|e\rangle$
are even under the exchange $1\leftrightarrow 2$, whereas $|{\bf t}\rangle$
and $|f_{-,\sigma}\rangle$ are odd. We ascribe positive parity also
to the operators $s^\dagger$, $f_{+}^\dagger$ and $e^\dagger$
and negative parity to ${\bf t}^\dagger$ and $f_{-}^\dagger$.
The procedure to transcribe operators for the original
t-J model to the representation in terms of bond particles
is as follows:
since the $9$ states introduced above form a complete basis of the
Fock space of a dimer, any operator $O$ acting within that dimer
can be expressed as $\sum_{a,b} |a \rangle O_{a,b} \langle b|$,
with $O_{a,b}=\langle a |O|b\rangle$.
Replacing $|a \rangle \langle b| \rightarrow a^\dagger b$ we obtain an
operator for the bond particles which has the same matrix elements
as long as it is acting in the subspace of states with precisely one
bond particle in the dimer.
In this way, the representation of the
spin operator at site $j \in \{1,2\}$ becomes
\begin{eqnarray}
{\bf S}_j &\rightarrow& \frac{\lambda_j}{2}\left(\;s^\dagger {\bf t}^{} + {\bf t}^\dagger s^{}\;\right)
-\frac{i}{2}\;{\bf t}^\dagger\times {\bf t}^{}\nonumber \\
&&\;\;\;\;\;\;\;\;+ \frac{1}{4}\;\left(\;{\bf f}_{+}^\dagger + \lambda_j {\bf f}_{-}^\dagger\;\right)
\bm{\tau} \left(\;{\bf f}_{+}^{} + \lambda_j {\bf f}_{-}^{}\;\right)
\label{spinop}
\end{eqnarray}
where $\lambda_j=(-1)^{j-1}$.
Introducing the contravariant spinor 
${\bf c}=(\hat{c}_\uparrow,\hat{c}_\downarrow)^T$
the representation of the electron annihilation operator at site 
$j$ becomes
\begin{eqnarray}
{\bf c}_j &\rightarrow& :\;\frac{1}{2}\;\left( s\;i\tau_y + 
\lambda_j {\bf t}\cdot\bm{\tau}i\tau_y\right)\;
\left(\;{\bf f}_{+}^\dagger - \lambda_j {\bf f}_{-}^\dagger\;\right)\nonumber \\
&&\;\;\;\;\;\;\;\;\;\;\;\;\;\;\;\;\;\;\;\;\;\;\;\;\;\;
+ \frac{1}{\sqrt{2}}\;e^\dagger \left(\;{\bf f}_{+}^{} + \lambda_j {\bf f}_{-}^{}\;\right)\;:
\label{cop}
\end{eqnarray}
where $:\dots :$ denotes normal ordering. 
As mentioned above, these representations are valid in the subspace of states 
with precisely one bond particle in the dimer:
\begin{eqnarray}
s^\dagger s^{} + {\bf t}^\dagger\cdot {\bf t}^{} +
\sum_\sigma\sum_{\alpha\in\{\pm\}} f_{\alpha,\sigma}^\dagger f_{\alpha,\sigma}^{}
+ e^\dagger e^{}= 1.
\label{constr}
\end{eqnarray}
\subsection{Generalization to the Plane}
\begin{figure}
\includegraphics[width=0.4\columnwidth]{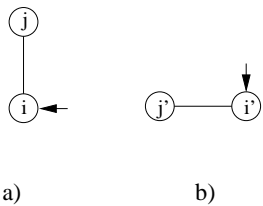}
\caption{\label{fig1} Under a rotation by $\pi/2$ in counterclockwise
direction the dimer in a) is transformed into the one in b).
The arrow indicates the marked site.}
\end{figure}

\begin{figure}
\includegraphics[width=0.35\columnwidth]{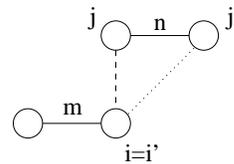}
\caption{\label{fig2} 
Hopping terms of different range (dashed lines) connect two dimers
in different ways.}
\end{figure}

We now consider the original plane with $N$ sites
and assume that these are partitioned into $\frac{N}{2}$ disjunct
dimers, whereby the sites in a dimer always are nearest neighbors. 
We call such a partitioning a dimer covering of the plane.
Each dimer is assigned a dimer label $m\in \{1,2,\dots ,N/2\}$.
We consider a dimer with dimer label $m$ which consists of
the sites $i$ and $j$.
When writing down the dimer states introduced above, we have to decide which of
the two sites,  $i$ or $j$, corresponds to the site $1$ and which one to
the site $2$ in (\ref{twoestates}) and (\ref{oneestates}). This is because
some of the dimer states have negative parity under $1\leftrightarrow 2$ 
so that their sign depends on this choice. We adopt 
the convention that for a bond in $x$-direction ($y$-direction)
the left (lower) site always corresponds to the site $1$.
We call the site which corresponds to $1$ the $1$-site and the
site which corresponds to $2$ the $2$-site of the dimer.
For each site $i$ we define $\lambda_i=1$ if it is the $1$-site
and $-1$ if it is the $2$-site in its respective dimer.
Next we introduce the bond particle  operators defined above and
give them an additional dimer label e.g. $s_{m}^\dagger$, 
${\bf t}_{m}^\dagger$ or $f_{m,+,\uparrow}^\dagger$. Lastly,
we define ${\bf R}_m=({\bf R}_i +{\bf R}_j)/2$.\\
Next we derive the bond-particle representation of the t-J Hamiltonian
for the given dimer covering. The intra-dimer part of the Hamiltonian is
\begin{eqnarray*}
H_{intra}&=&  \sum_m \left( - \frac{3J}{4} s_m^\dagger \;s_m^{}
+\frac{J}{4}\;{\bf t}_m^\dagger \cdot{\bf t}_m^{} \right.\nonumber \\
&& \left.
- t \sum_\sigma (\;f_{m,+,\sigma}^\dagger f_{m,+,\sigma}^{}
 -f_{m,-,\sigma}^\dagger f_{m,-,\sigma}^{}\;)\right).
\end{eqnarray*}
We proceed to the inter-dimer part of the Hamiltonian which
can can be constructed using (\ref{spinop}) and (\ref{cop}). 
Consider two dimers, $m$ and $n$, and let there be a bond
in the Hamiltonian which connects the sites $i$ and $j$
such that site $i$ belongs to dimer $m$, site $j$ to dimer $n$.
Using (\ref{spinop}) and (\ref{cop}) we then find the representations
\begin{widetext}
\begin{eqnarray}
J \;{\bf S}_{i}\cdot {\bf S}_{j} &\rightarrow&
\frac{J \lambda_i \lambda_{j}}{4}
\left(s_m^\dagger {\bf t}_m^{} + {\bf t}_m^\dagger s_m^{}\right)
\left(s_n^\dagger {\bf t}_n^{} + {\bf t}_n^\dagger s_n^{}\right)
-\frac{J}{4}\;({\bf t}_n^\dagger\times 
{\bf t}_n^{})\cdot({\bf t}_m^\dagger\times {\bf t}_m^{})
\nonumber \\
&&\;\;\;\;\;\;\;\;\;\;\;\;\;\;\;\;
-\frac{iJ}{4}\left[ \lambda_i \left(s_m^\dagger {\bf t}_m^{} + {\bf t}_m^\dagger s_m^{}\right)\cdot 
({\bf t}_n^\dagger\times {\bf t}_n^{})
+ \lambda_{j}
\left(s_n^\dagger {\bf t}_n^{} + {\bf t}_n^\dagger s_n^{}\right)
\cdot ({\bf t}_m^\dagger\times {\bf t}_m^{}) \right],
\label{ham1} \\[0.5cm]
-t \sum_\sigma \;\hat{c}_{i,\sigma}^\dagger \hat{c}_{j,\sigma}^{} &\rightarrow&
\frac{t}{4}\;[\;(\;s^\dagger_m s^{}_n\; + \lambda_i \lambda_{j} 
{\bf t}_m^\dagger \cdot {\bf t}_n^{}\;)\;\left(\;\sum_\sigma\; 
f_{n,j,\sigma}^\dagger f_{m,i,\sigma}^{}\;\right)\nonumber \\
&&\;\;\;\;\;\;\;\;\;\;\;\;\;\;\;\;
- (\;\lambda_i {\bf t}_m^\dagger s_n + \lambda_{j} s^\dagger_m {\bf t}_n^{}\;)
\cdot {\bf v}_{(n,j),(m,i)}
-i \lambda_i \lambda_{j} (\;{\bf t}_m^\dagger \times {\bf t}_n^{}\;)\cdot
{\bf v}_{(n,j),(m,i)}\;],
\label{ham2}
\end{eqnarray}
\end{widetext}
where 
\begin{eqnarray}
f_{m,i,\sigma}^{}&=&f_{m,+,\sigma}^{} -\lambda_i f_{m,-,\sigma}^{},
\label{fdef}
\end{eqnarray}
and the vector ${\bf v}$ is defined as
\begin{eqnarray*}
{\bf v}_{(n,j),(m,i)}&=& \sum_{\sigma,\sigma'}\;f_{n,j,\sigma}^\dagger\;
\bm{\tau}_{\sigma,\sigma'}\;f_{m,i,\sigma'}^{}.
\end{eqnarray*}
In (\ref{ham1}) and (\ref{ham2}) we have actually dropped all terms containing 
the boson $e_m^\dagger$ which represents the empty dimer, and also all
terms originating from the second line in (\ref{spinop}), which describes
direct exchange between holes. The reason is that both, the density of 
empty dimers and the contribution of direct exchange between holes
should be $\propto x^2$, and we consider $x$ to be small.\\
The factors of $\lambda_i$ or $\lambda_j$ in (\ref{ham1}) and (\ref{ham2})
guarantee the invariance of the Hamiltonian under symmetry operations
of the lattice. A given dimer covering of the plane could e.g. be
rotated by $\frac{\pi}{2}$ and the bond particle Hamiltonian for the
resulting dimer covering must be equivalent to the original one.
Consider a bond pointing in $y$-direction as in Figure \ref{fig1}a
which consists of the sites $i$ and $j$. According to our convention,
the site $i$ is the $1$-site so that the dimer state 
$|f_{-,\sigma}\rangle = (c_{i,\sigma}^\dagger - c_{j,\sigma}^\dagger)/\sqrt{2}$.
Now assume the whole dimer covering is rotated counterclockwise
by $\frac{\pi}{2}$. This is equivalent to a permutation of the numbers
of the lattice sites  and we assume that thereby $i\rightarrow i'$ and 
$j \rightarrow j'$. The transformed dimer state therefore is
$(c_{i',\sigma}^\dagger - c_{j',\sigma}^\dagger)/\sqrt{2}$, see  Figure \ref{fig1}b.
However, according to our convention, the $1$-site in the rotated dimer is
$j'$ so that the true dimer state is
$|f_{-,\sigma}\rangle= (c_{j',\sigma}^\dagger - c_{i',\sigma}^\dagger)/\sqrt{2}$.
The state $|f_{-,\sigma}\rangle$ thus aquires a factor of $(-1)$
under this operation and the same will hold true for any dimer state
with negative parity.
Now assume that one of the sites - say $i$ - in the dimer 
is `marked' and consider the product $\lambda_i|f_{-,\sigma}\rangle$.
The marked site in the rotated dimer then is $i'$. This is the $2$-site
in the rotated dimer so that $\lambda_{i'}=-1$ and the product 
$\lambda_i|f,-\rangle$ is invariant. This is easily seen to be general:
any operation which exchanges the $1$-site and the $2$-site in a dimer inverts 
both, the sign of a dimer state with odd parity and the sign of $\lambda$ for 
the marked site, so that their product stays invariant.
In (\ref{ham1}) and (\ref{ham2}) the marked sites
are the ones where the Hamiltonian acts and the factors of $\lambda$ are always
associated with dimer states of negative parity.\\
When the hopping term contains longer range hopping 
integrals such as $t'$ and $t''$, it may happen that two bonds
are connected by different hopping terms, as in Figure \ref{fig2}. 
We consider the factors of $\lambda$ in this case.
Let us assume the nearest neighbor term has $\lambda_i \lambda_{j}$ whereas 
the longer-range term has $\lambda_{i'} \lambda_{j'}$ as in Figure \ref{fig2}. 
Now consider a symmetry operation of the lattice. If the operation exchanges
the $1$-site and the $2$-site in bond $m$, 
both $\lambda_i$ and $\lambda_{i'}$ change sign. Whereas if the operation
exchanges the $1$-site and the $2$-site in bond $n$, both
$\lambda_{j}$ and  $\lambda_{j'}$ change sign.
It follows that the products $\lambda_i \lambda_{j}$ and 
$\lambda_{i'} \lambda_{j'}$ always change sign `in phase' so that they are 
equal up to an overall sign. We define $\xi=\pm 1$ as the relative sign of
the longer-range term with respect to the sign for nearest neighbor
hopping (or exchange): $\lambda_{i'} \lambda_{j'}=\xi \lambda_i \lambda_{j}$.
$\xi$ depends on the range of the hopping integral and
for $t'$ and $t''$ terms one finds $\xi=-1$. For any two sites
$i$ and $j$ we define $\xi_{i,j}$ to be this relative sign
for the hopping term which connects them.\\
\subsection{Approximations}
\begin{figure}
\includegraphics[width=0.8\columnwidth]{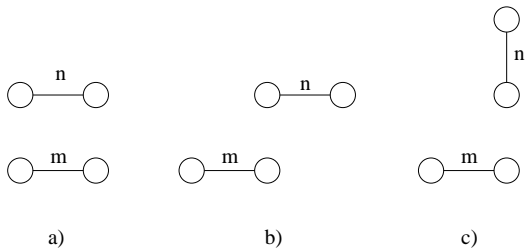}
\caption{\label{fig3} Estimation of the renormalization factor $\zeta$.}
\end{figure}

\begin{figure}
\includegraphics[width=0.2\columnwidth]{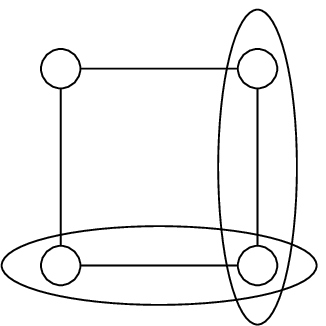}
\caption{\label{fig4} Two `crossing' bond particles - such configurations
are forbidden for the averaged Hamiltonian.}
\end{figure}

The representation of the problem in terms of dimer states is exact but
brings about no simplification so that approximations are necessary.
As a first step, we re-interpret the singlet as the vacuum 
state of a dimer and accordingly replace the corresponding operators 
$s_m^\dagger$ and $s_m^{}$ in (\ref{ham1}) and (\ref{ham2})
by unity. This is equivalent to the assumption that as
in a Mott-insulator at half-filling the electrons in the lightly doped 
Mott insulator form an inert background - the `singlet soup' - and that
the only remaining active degrees of freedom are the spins of the electrons -
represented by the triplets - and the doped holes, represented
by the fermions.
Replacing the singlet-operators in (\ref{ham1}) and (\ref{ham2}) by unity
we obtain terms of 2$^{nd}$, 3$^{rd}$ and 4$^{th}$ order in triplet 
and fermion operators. The constraint (\ref{constr})
to have precisely one bond particle per dimer then becomes
\begin{eqnarray}
{\bf t}_m^\dagger\cdot {\bf t}_m^{} +
\sum_{\alpha\in\{\pm\}}\sum_\sigma f_{m,\alpha,\sigma}^\dagger f_{m,\alpha,\sigma}^{}
+ e_m^\dagger e_m^{}\in\{0,1\},
\label{constr1}
\end{eqnarray}
which must hold for each bond $m$.
The constraint (\ref{constr1}) is equivalent to an infinitely strong repulsive 
potential between bond particles in the same dimer. For small $x$, however, 
the density of triplets and fermions is small - this will be seen below and 
is crucial for the present theory.
Namely for bose and fermi systems of low density it is known that even such
infinitely strong repulsive interactions between particles do not
qualitatively change the ground state\cite{Galitskii,Beliaev}. 
We therefore neglect the constraint assuming that it does not
lead to a qualitative change of the results due to the low density of
particles. \\
The problem with the infinitely strong repulsion between bond 
particles also occurs in the treatment of the Kondo lattice model.
There it was found that good qualitative agreement with numerical
results could be obtained by relaxing this constraint, and 
even quantitative agreement could be achieved by augmenting the energy of 
the bond particles by the loss of kinetic energy due to the blocking of bonds 
by a particle\cite{mykondo}. However, to keep the present treatment simple we 
do not introduce this correction here.
A more rigorous way to treat this repulsion has been carried out explicitely
for bond bosons in spin systems by Kotov {\em et al.}\cite{Kotov} and 
Shevchenko {\em et al.}\cite{Shevchenko}.\\
Even with the approximation to reduce the number of active degrees
of freedom we are far from a soluble problem, because the theory
still refers to a given dimer covering of the lattice so that it is
impossible to do any calculation for large systems. One might consider 
choosing a particular `simple' dimer covering.
However, since one is forced to make approximations, the exactness 
of the dimer representation is lost and the special symmetry of the
covering will make itself felt in the approximate solutions resulting e.g. in 
an artificial supercell structure.
On the other hand, the dimer Hamiltonian provides an exact representation of 
the original t-J model for {\em any} dimer covering of the plane. 
This means that for example the result for the spin correlation function 
$\langle {\bf S}_j(t)\cdot {\bf S}_i \rangle$ does not depend on the 
dimer covering in which the calculation is carried out. Put another way, the 
way in which a spin excitation  propagates through the network of dimers 
from site $i\rightarrow j$ during the time $t$ does not depend at all on the 
geometry of the dimer covering. This might suggest to construct a 
translationally invariant approximate Hamiltonian by {\em averaging} the dimer 
Hamiltonian over all possible coverings. This means that now every bond in the 
lattice may be occupied by a bond particle, and the averaged Hamiltonian for 
two bonds $n$ and $m$ is $\bar{h}_{n,m}=\zeta\;h_{n,m}$
where $h_{n,m}$ is given by the sum of (\ref{ham1}) and (\ref{ham2}) and
\begin{eqnarray}
\zeta=\frac{N_{n,m}}{N_d}.
\label{renorm}
\end{eqnarray}
Here $N_{n,m}$ is the 
number of dimer coverings which contain the bonds $n$ and $m$ and $N_d$ 
is the total number of dimer coverings. The resulting Hamiltonian obviously 
is translationally invariant and isotropic. To estimate $\zeta$ we use
a crude approximation: consider two adjacent bonds as in 
Figure \ref{fig3}. 
By symmetry the bond $m$ is covered by a dimer in exactly $1/4$ of all dimer
coverings and we restrict ourselves to these.
Assuming for simplicity that the number of coverings
containing one of the three possible orientations of the adjacent
bond $n$ are equal, we estimate  $\zeta=\frac{1}{12}$.
Later on, it will be seen that e.g. the spin gap depends sensitively on the 
value of $\zeta$ and we will consider it as an adjustable parameter but
the values which give `reasonable' results always
are around $\zeta=0.1$.\\
In the averaged Hamiltonian there are additional unphysical configurations.
For example, two bond particles may `cross' each other, see Figure \ref{fig4},
and such configurations have to be excluded as well.
This obviously amounts to an infinitely strong repulsion 
between the bond particles which acts whenever two bond particles share 
at least one site. Assuming the low-density limit we neglect this
repulsion.\\
Lastly, we discuss the operator of electron number. If all dimers in a given 
covering are occupied by singlets or triplets,
the number of electrons in the system is $N$. Each fermion reduces the number
of electrons by one so if we discard the $e^\dagger$-boson
\begin{eqnarray}
x &=&\frac{1}{N}\;\sum_{m,\sigma} \;\left(\;f_{m,+,\sigma}^\dagger f_{m,+,\sigma}^{}
+ f_{m,-,\sigma}^\dagger f_{m,-,\sigma}^{}\;\right).
\label{delcount}
\end{eqnarray}
Upon averaging we increase the number of bonds which can be
occupied by a particle from $N/2$ to $2N$. However, we retain
the expression (\ref{delcount}) which guarantees a fermi surface
with a volume propertional to the number of doped holes.\\
Lastly we mention the convention for the Fourier transform
of bond operators. 
Defining $\rho_{m,\alpha}$ (with $\alpha \in \{x,y\}$)
to be $1$ if $m$ is a bond in $\alpha$-direction and $0$ otherwise,
the Fourier transform of a bond operator is
(with $\alpha \in {x,y}$) 
\[
t_{{\bf k},\alpha}^\dagger
=\frac{1}{\sqrt{N}}\;
\sum_{m}\;\rho_{m,\alpha}\;e^{i {\bf k}  {\bf R}_m }\;t_{m}^\dagger.
\]
\section{Mean Field Theory}
\subsection{Mean-Field Decoupling}
We next treat the bond particle Hamiltonian in Hartree-Fock 
approximation, thereby assuming a ground state which is invariant under 
spin rotations\cite{SachdevBhatt,Gopalan}. Accordingly, we drop the 
terms in (\ref{ham1}) which are of 3$^{rd}$ order in the triplets, and the
terms in (\ref{ham2}) which contain a single triplet or the vector poduct of 
two triplets. After Hartree-Fock decoupling these terms would give 
expectation values such as $\langle {\bf t}_m \rangle$ or 
$\langle {\bf t}_m^\dagger\times{\bf t}_n^{} \rangle$ which vanish
in a spin-rotation invariant state.
For better clarity we give the remaining Hamiltonian
after all approximations discussed so far have been made.
The Hamiltonian is the sum of the following terms
\begin{eqnarray}
H_B^{(0)}&=&J\;\sum_m {\bf t}_m^\dagger\cdot{\bf t}_m^{} \nonumber \\
&& \;\;\;\;\;\;\;\;
+\frac{\zeta}{4}\sum_{m<n}\;\sum_{i\in m \atop j\in n}J_{i,j}\lambda_i\lambda_j
\;\left({\bf t}_m^\dagger\cdot{\bf t}_n^{} + H. c. \right)\nonumber \\
&& \;\;\;\;\;\;\;\;\;\;\;\;\;\;
+\frac{\zeta}{4}\sum_{m<n}\;\sum_{i\in m \atop j\in n}J_{i,j}\lambda_i\lambda_j
\;\left( {\bf t}_m^\dagger\cdot{\bf t}_n^\dagger
+ H. c.\right),\nonumber \\
H_F^{(0)}&=&-t \sum_{m,\sigma}\;\left(f_{m,+,\sigma}^\dagger f_{m,+,\sigma}^{}
- f_{m,-,\sigma}^\dagger f_{m,-,\sigma}^{}\right)\nonumber \\
&&\;\;\;\;\;\;\;\;\;\;\;\;\;\;\;\;+\frac{\zeta}{4}
\sum_{m,n}\;\sum_{i\in m \atop j\in n}
t_{i,j}\;
\sum_\sigma f_{n,j,\sigma}^\dagger f_{m,i,\sigma}^{},\nonumber \\
H_B^{(1)}&=&-\frac{\zeta}{4}\sum_{m<n}\;
\;\sum_{i\in m \atop j\in n} J_{i,j}\;
({\bf t}_n^\dagger\times{\bf t}_n^{})\cdot
({\bf t}_m^\dagger\times {\bf t}_m^{}),
\nonumber\\
H_{B,F}^{(1)}&=&\frac{\zeta}{4}\sum_{m,n}\;\sum_{i\in m \atop j\in n} 
t_{i,j}\lambda_i \lambda_{j} \;
{\bf t}_m^\dagger \cdot {\bf t}_n^{}
\;\sum_\sigma\; 
f_{n,j,\sigma}^\dagger f_{m,i,\sigma}^{} .\nonumber \\
\label{ham2a}
\end{eqnarray}
Here $J_{i,j}=J$ if $i$ and $j$ are nearest neighbors and zero otherwise.
$H_B^{(0)}$ and $H_F^{(0)}$ are the noninteracting parts for bosons and fermions,
$H_B^{(1)}$ describes the interaction between 
bosons, and $H_{B,F}^{(1)}$ the interaction between bosons and fermions.
The double cross product in $H_B^{(1)}$ can be rewritten as
\begin{eqnarray*}
({\bf t}_n^\dagger\times {\bf t}_n^{})\cdot({\bf t}_m^\dagger\times {\bf t}_m^{})
&=& \sum_{\lambda\ne \lambda'}\; \left( t_{m,\lambda}^\dagger  t_{n,\lambda}^\dagger\;
 t_{n,\lambda'}^{} t_{m,\lambda'}^{} -\right.
\nonumber \\
&&\;\;\;\;\;\;\;\;\;\;\;\;\;\;\;\;\left. t_{m,\lambda}^\dagger  t_{n,\lambda}^{} \;t_{n,\lambda'}^\dagger t_{m,\lambda'}^{}\right),
\end{eqnarray*}
where $\lambda,\lambda'\in\{x,y,z\}$ denote the spin of the triplet.
Upon mean-field factorization and using the spin-rotation invariance
of the ground state this becomes\cite{Gopalan}
\begin{eqnarray*}
\left(
\frac{2}{3}\;\langle {\bf t}_{m}^{}\cdot  {\bf t}_{n}^{}\rangle 
 \;{\bf t}_{m}^\dagger \cdot {\bf t}_{n}^\dagger -
\frac{2}{3}\;\langle {\bf t}_{m}^\dagger\cdot  {\bf t}_{n}^{}\rangle
 \;{\bf t}_{n}^\dagger\cdot  {\bf t}_{m}^{}\right) + H.c. 
\end{eqnarray*}
The expectation values 
of products of two triplet operators thereby take the form
\begin{eqnarray}
\langle {\bf t}_{m}^\dagger \cdot {\bf t}_{n}^{} \rangle &=& \lambda_i \;\lambda_{j}\;\theta_{m,n},\nonumber\\
\langle {\bf t}_{m}^{}\cdot  {\bf t}_{n}^{} \rangle &=& \lambda_i \;\lambda_{j}\;\eta_{m,n},
\label{restrictionb}
\end{eqnarray}
where $i\in m$ and $j\in n$ are the sites where the exchange term acts
and the `reduced expectation values' $\theta_{m,n}$ and
$\eta_{m,n}$ are identical for all symmetry equivalent
pairs of bonds $m$, $n$. This follows from the fact that e.g.
$(\lambda_i {\bf t}_{m}^{})\;(\lambda_{j}{\bf t}_{n}^{})$ is symmetry
invariant by construction.\\
In the mean-field factorization of $H_{B,F}^{(1)}$ we again encounter
the expectation values
$\langle {\bf t}_{m}^\dagger\cdot  {\bf t}_{n}^{}\rangle$ and in addition
fermionic expectation values. We define
\begin{eqnarray}
\chi_{m,n} &=& \sum_{i\in m \atop j\in n}
\xi_{i,j}\frac{t_{i,j}}{t}\;\sum_\sigma\;
\langle f_{m,i,\sigma}^\dagger f_{n,j,\sigma}^{} \rangle.
\label{restrictionf}
\end{eqnarray}
Thereby $\chi_{m,n}$ is the same for all symmetry equivalent
pairs of bonds $m$, $n$ which follows from the fact that
$f_{m,\nu,\sigma}^\dagger$ in (\ref{fdef}) is symmetry-invariant by construction.
The symmetry properties of $\theta_{m,n}$, $\eta_{m,n}$ and $\chi_{m,n}$
can also be verified by performing a Hartree-Fock calculation 
{\em without} imposing any symmetry properties of the expectation values - 
it turns out that the self-consistent expectation values always are
the same for all symmetry-equivalent bonds.\\
Upon decoupling the boson-fermion interaction term $H_{B,F}^{(1)}$
we obtain two separate problems, one for the bosons, the other
for the fermions. The bosonic mean-field Hamiltonian is
\begin{eqnarray}
H_{B}&&= J\sum_{m}\;{\bf t}_m^\dagger\cdot {\bf t}_m^{} +\nonumber \\
&& 
\zeta\;\sum_{\langle m,n\rangle}\left[\;\left(\;\Delta_{m,n}\;
{\bf t}_m^\dagger\cdot {\bf t}_n^\dagger
+ T_{m,n}\;{\bf t}_m^\dagger\cdot {\bf t}_n^{}\;\right)
+ H.c.\;\right],\nonumber \\
\label{mfham}
\end{eqnarray}
where $\langle m,n \rangle$ indicates a sum over all pairs
of bonds $m$ and $n$ connected by a nearest neighbor
bond of the Hamiltonian and
\begin{eqnarray}
\Delta_{m,n} &=&\lambda_i\lambda_{j}\;\;\;
\frac{J}{4}\left(1 - \frac{2}{3}\;\eta_{n,m} \right),\nonumber \\
T_{m,n} &=&\lambda_i\lambda_{j}\left[\,
\frac{J}{4}\left(1 + \frac{2}{3}\;\theta_{n,m} \right)
+ \frac{t}{4}\chi_{n,m,}\right].
\label{mfps}
\end{eqnarray}
Thereby $i\in m$ and $j\in n$ are the sites where the exchange term
acts - these are necessarily nearest neighbors.\\
The fermionic mean-field Hamiltonian is (omitting the spin index for
brevity)
\begin{eqnarray}
H_F &=& -t\;\sum_{m}\;\left(\;f_{m,+}^\dagger f_{m,+}^{}
- f_{m,-}^\dagger f_{m,-}^{}\;\right) \nonumber \\
&& \;\;\;\;\;\;\;\;\;\;\;\;\;\;\;\;\;
+ \zeta\sum_{m,n}\;\sum_{i\in m \atop j\in n}\;\tilde{T}_{m,n}^{i,j}
f_{m,i}^\dagger f_{n,j}^{} ,\nonumber \\
\tilde{T}_{m,n}^{i,j} &=& \frac{t_{i,j}}{4}\;(1 + \xi_{i,j}\theta_{n,m}).
\label{H_F}
\end{eqnarray}
Thereby $\tilde{T}_{m,n}^{i,j}$ depends only on the type of hopping term
($t$, $t'$ ot $t''$) which connects the sites $i$ and $j$ and is the same 
for all symmetry equivalent pairs of bonds $m$ and $n$.
Inserting (\ref{fdef}) we find
\begin{eqnarray*}
f_{m,i}^\dagger f_{n,j}^{} &=& f_{m,+}^\dagger f_{n,+}^{}
- \lambda_i f_{m,-}^\dagger f_{n,+}^{} 
\nonumber \\
&& \;\;\;\;\;\;\;\; 
-\lambda_{j} f_{m,+}^\dagger f_{n,-}^{} + \lambda_i \lambda_{j} f_{m,-}^\dagger f_{n,-}^{} .
\end{eqnarray*}
\subsection{The Bosonic Problem}
\begin{figure}
\includegraphics[width=0.9\columnwidth]{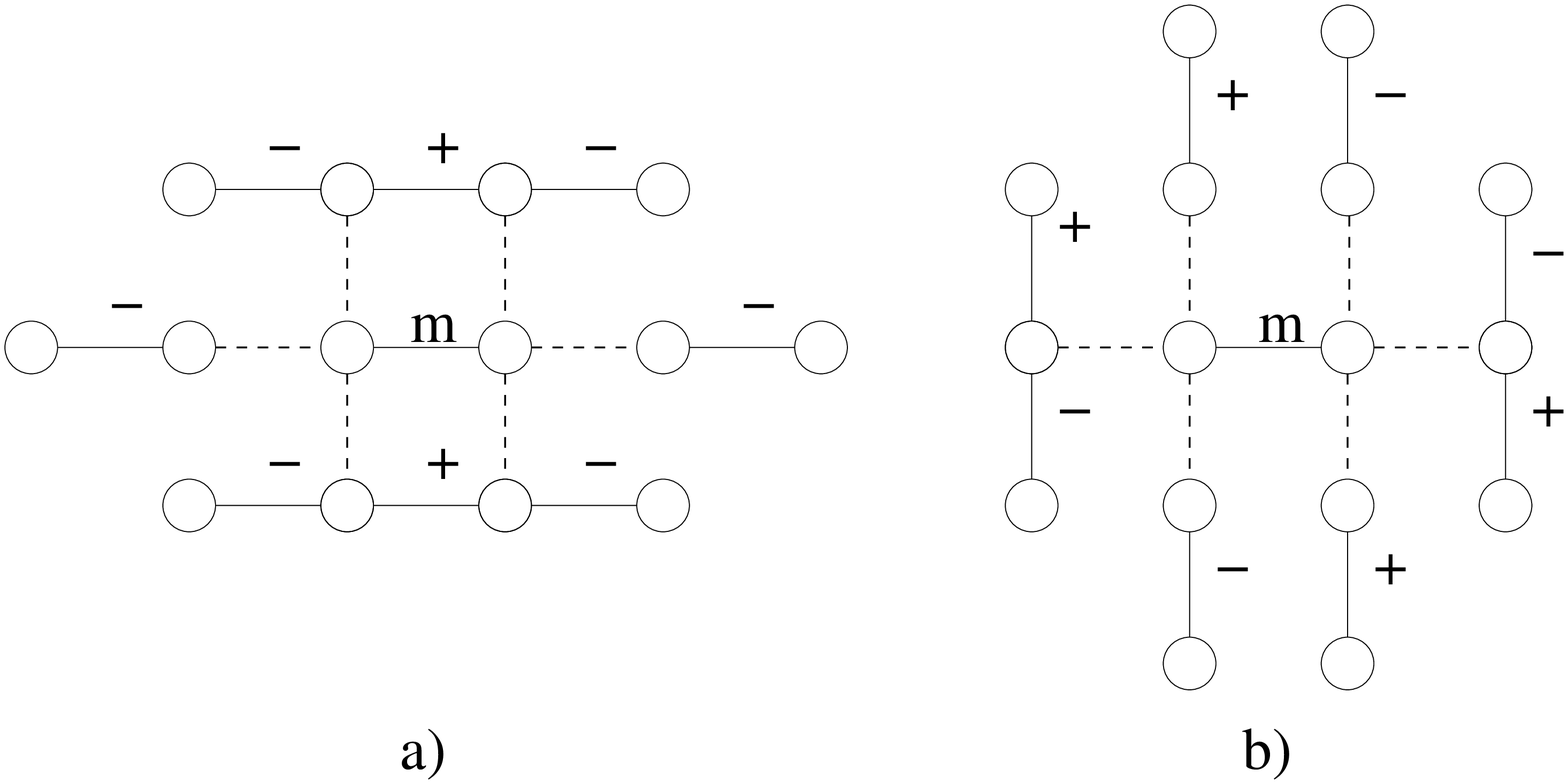}
\caption{\label{fig5} The factors of $\lambda_i\lambda_j$ for 
all bonds connected to the bond $m$ by the exchange term (dashed lines).
In a) both bonds are along the $x$-direction so that these
pairs contribute to $\tilde{\epsilon}^{\;x,x}$ whereas in
b) one bond is along $y$-direction so that these pairs contribute to
$\tilde{\epsilon}^{\;x,y}$.
In a) both bonds connecting parallel bonds have $\lambda_i\lambda_j=1$.}
\end{figure}
We consider $H_B$ in (\ref{mfham}).
The parameters $\theta_{m,n}$, $\eta_{m,n}$ and $\chi_{m,n}$
are identical for all symmetry-equivalent pairs of bonds. 
In the following
we replace e.g. $\theta_{m,n}\rightarrow \theta_{{\bf R}_m-{\bf R}_n}$ 
and the latter parameters are the same for all symmetry-equivalent
distances. Fourier transformation of $H_B$ then gives
\begin{eqnarray*}
H &=&\frac{1}{2}\;\sum_{\bf k} \;\sum_{\alpha,\beta\in\{x,y\}} \;\left( {\bf t}_{{\bf k},\alpha}^\dagger
\Delta^{\alpha,\beta}_{\bf k}\;{\bf t}_{\beta,-{\bf k}}^\dagger + H.c.\right)\\
&&\;\;\;\;\;\;\;\;\;\;\;\;\;\;\;\;\;\;\;\;
+ \sum_{\bf k}\;\;\sum_{\alpha,\beta\in\{x,y\}}{\bf t}_{{\bf k},\alpha}^\dagger\;
\epsilon^{\alpha,\beta}_{\bf k}\;{\bf t}_{\beta,{\bf k}}^{},
\end{eqnarray*}
with the $2\times 2$ matrices $\Delta_{\bf k}$ and $\epsilon_{\bf k}$ defined
by
\begin{eqnarray*}
\Delta^{\alpha,\beta}_{\bf k}&=&\frac{1}{N}\sum_{m,n} \;
\rho_{m,\alpha}\;\Delta_{m,n}\;\rho_{n,\beta}\;e^{i {\bf k}\cdot({\bf R}_n-{\bf R}_m)}
\end{eqnarray*}
and an analogous definition for $\epsilon^{\alpha,\beta}_{\bf k}$.
We assume $\Delta_{m,n}$ and $T_{m,n}$ to be real so that
$\epsilon_{-{\bf k}}=\epsilon^*_{\bf k}$ and
$\Delta_{-{\bf k}}=\Delta^*_{\bf k}$.
Both $\epsilon_{\bf k}$ and $\Delta_{\bf k}$ are Hermitean and
$\epsilon_{\bf k}=J + \zeta\;\tilde{\epsilon}_{\bf k}$ with
\begin{eqnarray*}
\tilde{\epsilon}^{\;x,x}_{\bf k}&=& 4T_{(0,1)}\cos(k_y) - 2T_{(2,0)}\cos(2k_x)\nonumber \\
&&\;\;\;\;\;\;\;\;\;\;\;\;\;\;\;\;\;\;\;\;\;- 4T_{(1,1)}\cos(k_x)\cos(k_y),\nonumber \\
&&\;\nonumber \\ 
\tilde{\epsilon}^{\;y,y}_{\bf k}&=& 4T_{(1,0)}\cos(k_x) - 2T_{(0,2)}\cos(2k_y)\nonumber \\
&&\;\;\;\;\;\;\;\;\;\;\;\;\;\;\;\;\;\;\;\;\;- 4T_{(1,1)}\cos(k_x)\cos(k_y),\nonumber \\
&&\;\nonumber \\ 
\tilde{\epsilon}^{\;x,y}_{\bf k}&=& 4 T_{(\frac{3}{2},\frac{1}{2})}
\left(\;\sin(\frac{3k_x}{2})\sin(\frac{k_y}{2}) \right. \nonumber \\
&&\;\;\;\;\;\;\;\;\;\;\;\;\;\;\;\;\;\;\;\;\;\left.
+\sin(\frac{k_x}{2})\sin(\frac{3k_y}{2})\;\right).
\end{eqnarray*}
This can be verified using the signs of the products
$\lambda_i\lambda_j$ in Figure \ref{fig5}.
The expressions for $\Delta_{\bf k}$ are obtained from
those for $\tilde{\epsilon}_{\bf k}$ by replacing
$T_{{\bf R}}\rightarrow \Delta_{{\bf R}}$. To diagonalize $H$ we make the ansatz
\begin{eqnarray}
{\bm \tau}_{\nu,{\bf k}}^\dagger \;\;\,&=&
\sum_{\alpha\in\{x,y\}}\;\left(\;u_{\nu,{\bf k},\alpha}\;{\bf t}_{{\bf k},\alpha}^\dagger
+ v_{\nu,{\bf k},\alpha}\;{\bf t}_{-{\bf k},\alpha}^{}\;\right),\nonumber \\
{\bm \tau}_{\nu,-{\bf k}}^{} &=&
\sum_{\alpha\in\{x,y\}}\;\left(\;v^*_{\nu,{\bf k},\alpha}\;{\bf t}_{{\bf k},\alpha}^\dagger
+ u^*_{\nu,{\bf k},\alpha}\;{\bf t}_{-{\bf k},\alpha}^{}\;\right),
\label{bogo}
\end{eqnarray}
where $\nu \in \{1,2\}$. 
Demanding $[\tau_{\nu,{\bf k}}^{},\tau_{\mu,{\bf k}}^\dagger]=
\delta_{\nu,\mu}$
leads to
\begin{eqnarray}
\sum_{\alpha\in\{x,y\}}\;\left(\;u^*_{\nu,{\bf k},\alpha}u^{}_{\mu,{\bf k},\alpha} 
- v^*_{\nu,{\bf k},\alpha}v^{}_{\mu,{\bf k},\alpha}\;\right) &=& \delta_{\nu,\mu}.
\label{norm}
\end{eqnarray}
whereas $[H,{\bm \tau}_{\nu,{\bf k}}^\dagger]=\omega_{\nu,{\bf k}}{\bm \tau}_{\nu,{\bf k}}^\dagger$
results in the non-Hermitean eigenvalue problem
\begin{eqnarray}
\left(\begin{array}{c c}
\epsilon_{\bf k}   & -\Delta_{\bf k}\\
\Delta^*_{-{\bf k}} & -\epsilon^*_{-{\bf k}}
\end{array} \right)\;
\left(\begin{array}{c}
u_{\nu,{\bm k}} \\
v_{\nu,{\bm k}}
\end{array} \right) &=& \omega_{\nu,{\bf k}}\;
\left(\begin{array}{c}
u_{\nu,{\bm k}} \\
v_{\nu,{\bm k}}
\end{array} \right).
\label{mats}
\end{eqnarray}
For a matrix of the type on the left hand side 
it can be shown that the eigenvalues come in pairs of $\pm \omega$ 
and if $(u,v)$ is the right eigenvector for $+\omega$ then
$(v^*,u^*)$ is the right eigenvector for $-\omega$ - which justifies the 
ansatz (\ref{bogo}). Moreover $(u^*,-v^*)$ can be shown
to be the left eigenvector for $+\omega$
so that (\ref{norm}) is equivalent to the condition that left and right 
eigenvectors for different eigenvalues are orthogonal - as it has to be.
The mean-field Hamiltonian becomes
\begin{eqnarray*}
H_B &=& \sum_{\bf k}\left(\sum_{\nu=1}^2\;\omega_{\nu,{\bf k}}
\left({\bm {\bm \tau}}_{\nu,{\bf k}}^\dagger\;{\bm {\bm \tau}}_{\nu,{\bf k}}^{} + \frac{3}{2}\right)
- \frac{3}{2}\;tr(\epsilon_{\bf k})\right),
\end{eqnarray*}
and (\ref{bogo}) can be reverted to give
\begin{eqnarray*}
{\bf t}_{{\bf k},\alpha}^\dagger\;\;\, &=& \sum_{\nu=1}^2\;\left(\;\;\;\,
u_{\nu,{\bf k},\alpha}^*\;{\bm \tau}_{{\bf k},\nu}^\dagger - v_{\nu,{\bf k},\alpha}\;{\bm \tau}_{-{\bf k},\nu}^{}
\;\right),\nonumber \\
{\bf t}_{-{\bf k},\alpha}^{} &=& \sum_{\nu=1}^2\;\left(\;
-v_{\nu,{\bf k},\alpha}^*\;{\bm \tau}_{{\bf k},\nu}^\dagger + 
u_{\nu,{\bf k},\alpha}\;{\bm \tau}_{-{\bf k},\nu}^{}\;\right).
\end{eqnarray*}
Using these expressions, the expectation values
$\langle {\bf t}_{{\bf k},\alpha}^\dagger\cdot  {\bf t}_{{\bf k},\beta}^{}  \rangle$
and $\langle {\bf t}_{{\bf k},\alpha}^{} \cdot{\bf t}_{-{\bf k},\beta}^{} \rangle$
can be obtained, from which the parameters
$\theta_{m,n}$ and $\eta_{m,n}$ in (\ref{restrictionb}) can be calculated.\\
\subsection{The Fermionic Problem}
We consider the fermionic Hamiltonian (\ref{H_F}).
Fourier transformation gives $H_F=\sum_{\bf k}H_{F,{\bf k}}$ 
\begin{eqnarray}
H_{F,{\bf k}} &=&-t\sum_{\alpha\in\{x,y\}}
\left(f_{{\bf k},\alpha,+}^\dagger f_{{\bf k},\alpha,+}^{}
- f_{{\bf k},\alpha,-}^\dagger f_{{\bf k},\alpha,-}^{}\right) \nonumber \\
&&\;\;\;\;\;\;\;\;\;\;\;\;\;\;\;\;\;\;\;\;\;\;\;\;\;\;\;\;\;\;\;\;\;\;\;
+ \zeta\;v_{\bf k}^\dagger\tilde{H}_{\bf k}^{}v_{\bf k}^{}
\label{HFK}
\end{eqnarray}
where the vector $v_{\bf k}=(f_{k,x,+},  f_{k,y,+}, f_{k,x,-},  f_{k,y,-})^T$
has been introduced.
Here we give the elements of the $4\times 4$ matrix $\tilde{H}_{\bf k}$ for 
the case that
the t-J Hamiltonian contains only nearest neighbor hopping:
\begin{widetext}
\begin{eqnarray*}
      \tilde{H}_{1,1}&=& 4 \tilde{T}_{(0,1)} \cos(k_y) + 4 \tilde{T}_{(1,1)} \cos(k_x) \cos(k_y) + 2 \tilde{T}_{(2,0)} \cos(2 k_x)\\
      \tilde{H}_{1,2}&=& 4   \left(\; \tilde{T}_{(\frac{3}{2},\frac{1}{2})} \cos(\frac{3k_x}{2}) \cos(\frac{k_y}{2}) + \tilde{T}_{(\frac{1}{2},\frac{3}{2})} \cos(\frac{k_x}{2}) \cos(\frac{3k_y}{2})\;\right)\\
      \tilde{H}_{1,3}&=& - 2   i   \left(\; 2 \tilde{T}_{(1,1)} \sin(k_x) \cos(k_y) + \tilde{T}_{(2,0)} \sin(2 k_x) \;\right)\\
      \tilde{H}_{1,4}&=& - 4   i   \left(\; \; \tilde{T}_{(\frac{3}{2},\frac{1}{2})} \cos(\frac{3k_x}{2}) \sin(\frac{k_y}{2}) + \tilde{T}_{(\frac{1}{2},\frac{3}{2})} \cos(\frac{k_x}{2}) \sin(\frac{3k_y}{2}) \;\right)\\
      \tilde{H}_{2,2}&=& 4 \tilde{T}_{(1,0)} \cos(k_x) + 4 \tilde{T}_{(1,1)} \cos(k_x) \cos(k_y) + 2 \tilde{T}_{(0,2)} \cos(2 k_y)\\
      \tilde{H}_{2,3}&=& - 4 i \left(\; \tilde{T}_{(\frac{3}{2},\frac{1}{2})} \sin(\frac{3k_x}{2}) \cos(\frac{k_y}{2}) + \tilde{T}_{(\frac{1}{2},\frac{3}{2})} \sin(\frac{k_x}{2}) \cos(\frac{3k_y}{2}) \;\right)\\
      \tilde{H}_{2,4}&=& - 2 i  \left(\; 2 \tilde{T}_{(1,1)} \cos(k_x) \sin(k_y)  + \tilde{T}_{(0,2)} \sin(2 k_y) \;\right)\\
      \tilde{H}_{3,3}&=& 4 \tilde{T}_{(0,1)} \cos(k_y) - 4 \tilde{T}_{(1,1)} \cos(k_x) \cos(k_y) - 2 \tilde{T}_{(2,0)} \cos(2 k_x)\\
\end{eqnarray*}
\begin{eqnarray*}
      \tilde{H}_{3,4}&=& 4 \left(\; \tilde{T}_{(\frac{3}{2},\frac{1}{2})} \sin(\frac{3k_x}{2}) \sin(\frac{k_y}{2}) + \tilde{T}_{(\frac{1}{2},\frac{3}{2})} \sin(\frac{k_x}{2}) \sin(\frac{3k_y}{2}) \;\right)\\
      \tilde{H}_{4,4}&=& 4 \tilde{T}_{(1,0)} \cos(k_x) - 4 \tilde{T}_{(1,1)} \cos(k_x) \cos(k_y) - 2 \tilde{T}_{(0,2)} \cos(2 k_y)\\
\end{eqnarray*} 
\end{widetext}
Note that on the respective right-hand side we have used the
notation $\tilde{T}_{m,n}\rightarrow \tilde{T}_{{\bf R}_m-\bf{R}_n}$ and dropped
the superscripts on $\tilde{T}$ because they all refer to nearest neighbors.
The terms in $\tilde{H}$
originating from longer range hopping integrals such as $t'$ and 
$t''$ are also easily written down but the resulting expressions are lengthy
so we do not give them here.\\
Diagonalizing $-(H_{F,{\bf k}}-\mu)$ we obtain the dispersion
$E_{\nu,{\bf k}}$ (with $\nu=1\dots 4$)
for the electron-like quasiparticles and the corresponding
eigenvectors ${\bf e}_{\nu,{\bf k}}$, from which the mean-field expectation 
values $\chi_{n,m}$ (\ref{restrictionf}) can be calculated. 
This allows to perform the self-consistency procedure, thereby using 
Broyden's algorithm\cite{Broyden} for better convergence, and obtain the 
self-consistent values of the bosonic parameters $\theta_{m,n}$ and $\eta_{m,n}$  
in (\ref{restrictionb}) and $\chi_{n,m}$ (\ref{restrictionf}).
\subsection{Excitation Spectra}
Having obtained a self-consistent solution we can evaluate physical
properties such as the single-particle spectral function and the
dynamic spin structure factor. We consider a dimer $m$, which comprises
the sites $i$ and $j$. Using (\ref{spinop}) and (\ref{cop}) 
one finds the following representations\cite{Gopalan}
\begin{widetext}
\begin{eqnarray}
\sum_{\nu\in\{i,j\}}\;e^{i {\bm k}\cdot {\bm R}_\nu}\;\hat{c}_{\nu}
&\rightarrow&e^{i {\bm k}\cdot{\bm R}_m}\;
i\tau_y\;\left(\;
\cos\left(\frac{k_\alpha}{2}\right)\;f_{m,+}^\dagger + 
i\sin\left(\frac{k_\alpha}{2}\right)\;f_{m,-}^\dagger\;
\right)\nonumber \\
&&\;\;\;\;\;\;\;\;\;\;\;\;\;\;\;\;\;\;\;\;
- {\bm t}_m\cdot{\bm \tau}i\tau_y\;\left(\;
\cos\left(\frac{k_\alpha}{2}\right)\;f_{m,-}^\dagger 
+ i\sin\left(\frac{k_\alpha}{2}\right)\;f_{m,+}^\dagger\;
\right),
\nonumber \\
\sum_{\nu\in\{i,j\}}\;e^{i {\bm k}\cdot {\bm R}_\nu}\;
{\bm S}_{\nu}
&\rightarrow&-i\;e^{i {\bm k}\cdot{\bm R}_m}\;
\;\left(\sin\left(\frac{q_\alpha}{2}\right)({\bm t}_m^\dagger +{\bm t}_m^{})
+ \cos\left(\frac{q_\alpha}{2}\right)\;{\bm t}_m^\dagger \times
{\bm t}_m^{}\;\right),
\label{oprep}
\end{eqnarray}
\end{widetext}
where $\alpha\in \{x,y\}$ is the direction of the bond. 
Upon Fourier transformation, the respective second terms in both equations 
describe processes where a triplet which is already present in the system
is annihilated - since the momentum of the triplet is not fixed, this
would result in incoherent continua. Accordingly, we drop these terms
and retain only the first terms in both equations, which give 
$\delta$-peaks. We have for any operator $O_i$
\begin{eqnarray*}
\frac{1}{N}\sum_i e^{i{\bf k}\cdot {\bf R}_i} O_i =
\frac{1}{4N}\sum_m \left(\;\sum_{i\in m} e^{i{\bf k}\cdot {\bf R}_i} O_i\right)
\end{eqnarray*}
where $\sum_m$ denotes a sum over all nearest neighbor bonds
and inserting (\ref{oprep}) for the bracket on the right-hand side
\begin{eqnarray*}
\hat{c}_{-{\bm k},\uparrow}&=& 
\frac{1}{4}\;\sum_{\alpha\in\{x,y\}}\left(
\cos\left(\frac{k_\alpha}{2}\right)f_{{\bf k},\alpha,+,\uparrow}^\dagger \right.
\nonumber \\
&& \;\;\;\;\;\;\;\;\;\;\;\;\;\;\;\;\;\;\;\;\;\;\;\;\left. 
+ i\sin\left(\frac{k_\alpha}{2}\right)
f_{{\bf k},\alpha,-,\uparrow}^\dagger\right),\nonumber 
\end{eqnarray*}
\begin{eqnarray*}
S_{\bf q}&=&-\frac{i}{4}\sum_{\alpha\in\{x,y\}} \sin\left(\frac{q_\alpha}{2}\right)
({\bf t}_{{\bf q},\alpha}^\dagger + {\bf t}_{-{\bf q},\alpha}^{}).
\end{eqnarray*}
We then obtain the single-particle spectral function and the dynamic
spin structure factor
\begin{eqnarray}
A({\bf k},\omega)&=&\sum_{\nu=1}^4\; | e^*_{\nu,{\bf k}} \cdot a_{\bf k}|^2
\delta(\omega - E_{\nu,{\bf k}}) \label{akw} \\
S({\bf q},\omega) &=& \int d\omega \;e^{i\omega t}
\langle S_{-{\bm q}}(t) S_{{\bm q}} \rangle \nonumber \\
&=& \sum_{\nu=1}^2\; |F_\nu({\bm q})|^2
\delta(\omega - \omega_{\nu,{\bf q}}) \nonumber \\
F_\nu({\bm q})&=&\sum_{\alpha\in\{x,y\}}\sin\left(\frac{q_\alpha}{2}\right)
(u_{\nu,{\bf q},\alpha}^* - v_{\nu,{\bf q},\alpha}^*)
\label{sqw}
\end{eqnarray}
where $a_{\bf k}=\left(\cos\left(\frac{k_x}{2}\right),
\cos\left(\frac{k_y}{2}\right),
i\sin\left(\frac{k_x}{2}\right),i\sin\left(\frac{k_y}{2}\right)\right)^T$.
Using these expressions the coherent spectral weight of the individual bands 
in the electron spectral function and spin structure factor
can be calculated.
\section{Results}
\begin{figure}
\includegraphics[width=0.6\columnwidth]{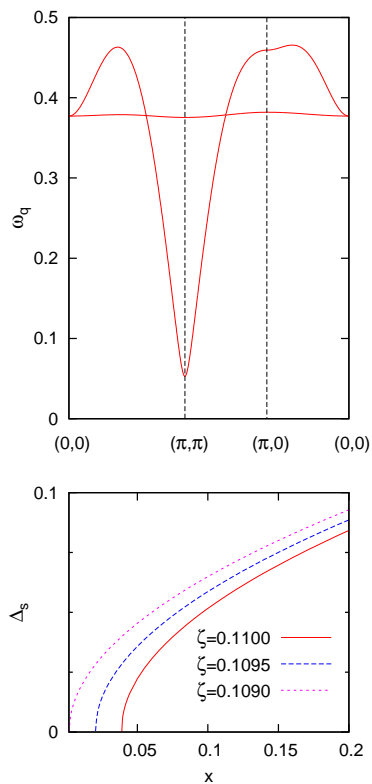}
\caption{\label{fig6} Top: Triplet dispersion $\omega_{\nu,{\bm q}}$
(see (\ref{mats})). Parameter values are $J=0.4$, $t'=t''=0$, $x=0.1$, and 
$\zeta=0.11$.
Bottom: Spin gap $\Delta_s=\omega_{(\pi,\pi)}$ versus $x$
for different $\zeta$, $J=0.4$, $t'=t''=0$.}
\end{figure}

Performing the self-consistency procedure described in the
preceding section gives the self-consistent values of the mean-field parameters
$\theta$, $\eta$ and $\chi$. 
Using these, the triplet frequencies 
$\omega_{\nu,{\bf q}}$ (with $\nu\in\{1,2\}$), the energies of the electron-like 
quasiparicles $E_{\nu,{\bf k}}$ (with $\nu\in\{1\dots 4\}$) and the excitation 
spectra can be calculated. The results presented below
were obtained at inverse temperature $\beta=200$.\\
Figure \ref{fig6} shows the dispersion $\omega_{\nu,{\bf q}}$
for the the triplet bosons. There is a dispersive band
with minimum at $(\pi,\pi)$ and a second band with practically no
dispersion.  The coherent spectral weight $|F_\nu({\bm q})|^2$ in the
spin correlation function (\ref{sqw}) vanishes for this band.
Most likely this band therefore is an artefact of
the enlargement of the basis of dimer particles in the course of the
averaging procedure  - an obvious drawback of the present approximation.
The lower part of Figure \ref{fig6} shows the energy of the dispersive
band at $(\pi,\pi)$ - which we call the spin gap $\Delta_S$ - as a function of
$x$. This is shown for different values of the parameter
$\zeta$ which originates in the averaging procedure.
For larger $x$, $\Delta_S$ shows a roughly linear variation with 
$x$ but bends down sharply as $x\rightarrow 0$ and reaches zero
at a certain $x$ which sensitively depends on $\zeta$.
It is a plausible scenario that $\Delta_S\rightarrow 0$ for some small 
$x$, so that the triplets condense into momentum $(\pi,\pi)$ resulting 
in antiferromagnetic order\cite{SachdevBhatt}. In this case, the triplet 
dispersion would be backfolded, resulting in a dispersion that is 
quite similar to that of antiferromagnetic magnons. The bandwidth, however, is 
only  $\approx J$ whereas it should be $2J$, another 
deficiency of the present approximation. Since the precise value of $x$ where 
antiferromagnetic
order sets in is unknown we fix $\zeta=0.11$ from now on.\\
\begin{table}
\begin{center}
\begin{tabular}{|c|r r|}
\hline
 & A & B\\
\hline
 $\theta_{(1,0)}$                 &  -0.0234 &  -0.0233 \\ 
 $\theta_{(1,1)}$                 &  -0.0198 &  -0.0196 \\   
 $\theta_{(2,0)}$                 &  -0.0229 &  -0.0228 \\   
 $\theta_{(\frac{3}{2},\frac{1}{2})}$ &  -0.0223 &  -0.0221 \\ 
\hline
 $\eta_{(1,0)}$                  &  -0.1178 &  -0.1176 \\   
 $\eta_{(1,1)}$                  &  -0.0673 &  -0.0672 \\   
 $\eta_{(2,0)}$                  &  -0.0720 &  -0.0718 \\   
 $\eta_{(\frac{3}{2},\frac{1}{2})}$  &  -0.0712  & -0.0710 \\ 
\hline
 $\chi_{(1,0)}$                  &  -0.0330 &  -0.0210 \\   
 $\chi_{(1,1)}$                  &  -0.0347 &  -0.0356 \\   
 $\chi_{(2,0)}$                  &  -0.0110 &   0.0013 \\   
 $\chi_{(\frac{3}{2},\frac{1}{2})}$  &  -0.0150 &  -0.0283 \\ 
\hline
\end{tabular}
\caption{Self-consistent mean-field expectation values
$\theta_{\bf R}$ and $\eta_{\bf R}$ in (\ref{restrictionb})
and $\chi_{\bf R}$ in (\ref{restrictionf}),
for $J=0.4$, $x=0.1$ and 
$\zeta=0.11$ whereby $t'=t''=0$ (A) and $t'=-0.2$ $t''=0.1$ (B).}
\label{tab1}
\end{center}
\end{table}
Table \ref{tab1} gives the values of the self-consistent parameters
$\theta_{\bf R}$, $\eta_{\bf R}$ and $\chi_{\bf R}$. These are small
so that for any approximate calculation at finite-doping - where the vanishing
of $\Delta_S$ is of no concern - the self-consistent parameters also
could be simply omitted. This was noted previously by Gopalan {\em et al.}
in their study of spin-ladders\cite{Gopalan}.
\begin{figure}
\includegraphics[width=\columnwidth]{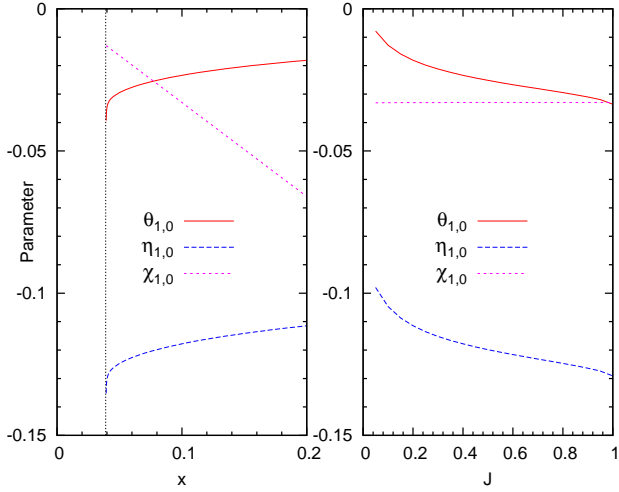}
\caption{\label{fig7} Variation of mean-field parameters
with $x$ (for $J=0.4$) and with $J$ (for $x=0.1$). The vertical line denotes
the $x$ where $\Delta_S\rightarrow 0$. Other parameter values are
$t'=t''=0$, $\zeta=0.11$.
}
\end{figure}
Figure \ref{fig7} shows some self-consistent mean-field parameters
as a function of $x$ and $J$. Except for
a small range near the critical $x$ where the spin gap closes
the bosonic parameters $\theta$ and $\eta$ show little variation with either 
$x$ or $J$. The fermionic parameter $\chi$ is linear in $x$ as expected, and 
practically
independent of $J$. Figure \ref{fig8} shows the density of bosons per
bond, $n_B$, and the combined density of bosons and fermions per bond,
$n_B+\frac{x}{2}$, versus $x$. As already mentioned the densities
are small so that relaxing the various constraints on the bond particles 
may be reasonably justified.
\begin{figure}
\includegraphics[width=0.6\columnwidth]{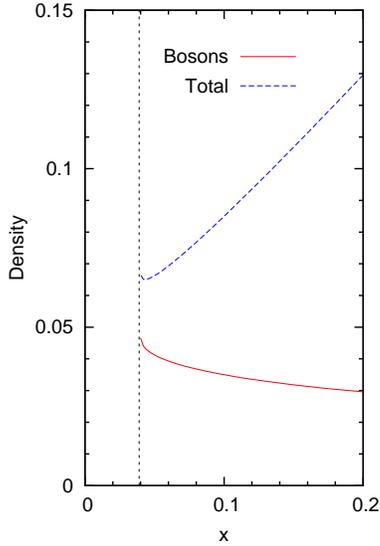}
\caption{\label{fig8} Density of bosons and density of
bosons and fermions per bond versus hole concentration $x$.
The vertical line denotes the $x$ where $\Delta_S\rightarrow 0$.
Parameter values are $J=0.4$, $t'=t''=0$, and $\zeta=0.11$.
}
\end{figure}
Figure \ref{fig9} shows the band structure for electron-like
quasiparticles, $E_{\nu,{\bf k}}$.
The topmost band has a maximum between $(\frac{\pi}{2},\frac{\pi}{2})$
and $(\pi,\pi)$. For vanishing $t'$ and $t''$
this maximum actually is degenerate along a circular
contour around $(\pi,\pi)$, so that the fermi surface for finite
$x$ would be a ring with tiny width aroud $(\pi,\pi)$ 
(the area covered by the ring would be a fraction $x/2$ of the total Brillouin
zone). This is obviously unphysical. In addition there are two
dispersionless bands at $\approx - 0.25 t$ and $\approx - 2.14 t$.
As was the case for the dispersionless band in the triplet dispersion,
these bands have zero coherent weight in the single particle
spectral function so we again interpret them as being artefacts of the 
averaging procedure.\\
\begin{figure}
\includegraphics[width=0.6\columnwidth]{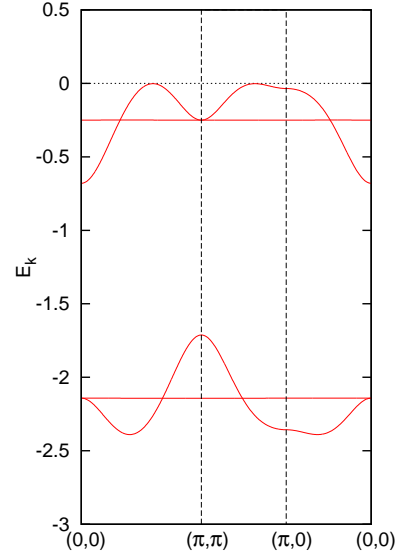}
\caption{\label{fig9} Quasiparticle dispersion $E_{\nu,{\bf k}}$
(obtained by diagonalizing $-(H_{F,{\bf k}}-\mu)$ with $H_{F,{\bf k}}$
in (\ref{HFK})). Parameter values are $J=0.4$, $t'=t''=0$, $x=0.1$, 
and $\zeta=0.11$.}
\end{figure}
From now on we consider the system with additional longer range hopping
integrals $t'=-0.2$, $t''=0.1$ and $x=0.1$. The self-consistent mean-field 
parameters for this case are also given in Table \ref{tab1}.
Figure \ref{fig10} shows the band structure $E_{\nu,{\bm k}}$ and the
single-particle spectral density $A({\bm k},\omega)$,
\begin{figure}
\includegraphics[width=0.8\columnwidth]{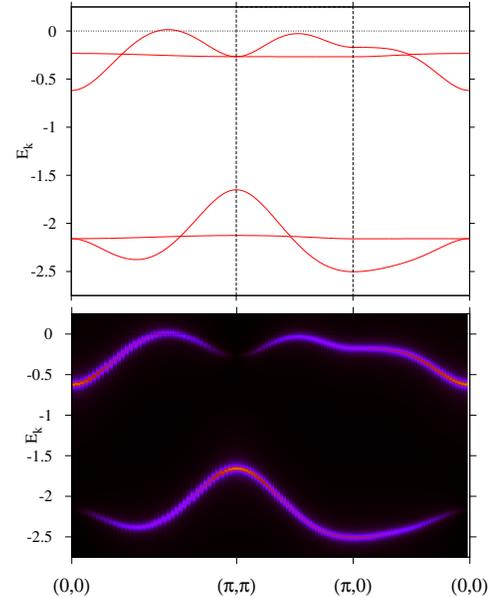}
\caption{\label{fig10} Top: Quasiparticle dispersion $E_{\nu,{\bm k}}$
for $J=0.4$, $t'=-0.2$, $t''=0.1$, $x=0.1$ and $\zeta=0.11$.
Bottom: Corresponding spectral density $A({\bm k},\omega)$,
see Eq. (\ref{akw}). $\delta$-functions have been replaced by
Lorentzians of with $\varepsilon=0.02$.}
\end{figure}
Figure \ref{fig11} the fermi surface.
Figure \ref{fig10} confirms that the flat bands
have no spectral weight - in fact the spectral weight of these bands is not 
just small but zero to computer accuracy. Qualitatively, the topmost 
dispersive band 
which crosses $\mu$ can be compared to ARPES results in several aspects. 
Its maximum is at $(0.59\pi,0.59\pi)$, so 
that the fermi surface is a hole pocket centered at this point - see Figure
\ref{fig11}. The spectral weight of this band decreases as one moves towards 
$(\pi,\pi)$ so that the outer edge of the pocket has a smaller spectral weight.
However, the hole pocket is too close to $(\pi,\pi)$ and the drop of spectral 
weight is far from being steep enough to really match experiment.
\begin{figure}
\includegraphics[width=0.6\columnwidth]{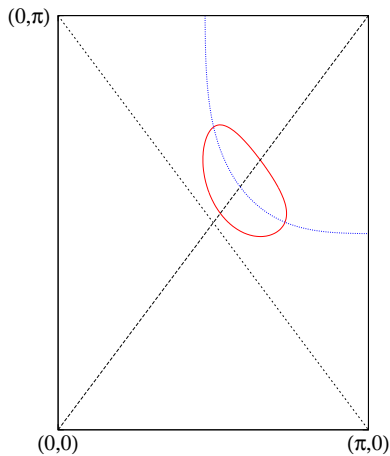}
\caption{\label{fig11} Fermi surface for the parameter values
in Fig. \ref{fig10} (red). The blue line marks the maximum of the band
as a function of the angle with respect to the line $(0,\pi)-(\pi,\pi)$.}
\end{figure}
Along $(\pi,0)\rightarrow(\pi,\pi)$ the band first disperses towards $\mu$.
then bends down and looses weight around the bending point.
This is qualitatively similar as in experiment\cite{Hashimoto} but the bending
point $(\pi,0.47\pi)$ is too far from $(\pi,0)$, the band is too far 
from $\mu$ at $(\pi,0)$ and the drop of spectral is much too smooth. 
On the other hand, this is only the mean-field result and coupling to the 
triplet bosons may lead to modifications of the quasiparticle dispersion
and spectral weight, as is the case for hole motion in an 
antiferromagnet\cite{Bulaevskii,Trugman,Becker,MartinezHorsch,ChenSushov}.\\
\begin{figure}
\includegraphics[width=0.8\columnwidth]{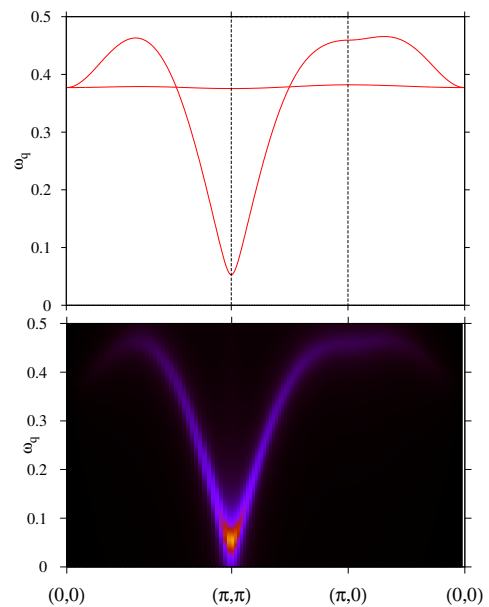}
\caption{\label{fig12} Top: Triplet dispersion $\omega_{\nu,{\bm q}}$
(see (\ref{mats})) for the parameter values in Figure \ref{fig10}.
Bottom: Coherent spectral weight in the dynamic spin structure factor, 
see Eq. (\ref{sqw}). $\delta$-functions have been replaced by
Lorentzians of with $\varepsilon=0.02$.}
\end{figure}
Figure \ref{fig12} shows the triplet dispersion $\omega_{\nu,{\bm q}}$
and the spectral intensity in the spin correlation function $S({\bf q},\omega)$
The dispersionless band has zero spectral weight, so that only a single mode
is visible in $S({\bf q},\omega)$. This has a minimum
at $(\pi,\pi)$ and the spectral weight is concentrated sharply around
this wave vector. Experimentally, inelastic neutron scattering from underdoped 
cuprates shows an `hourglass' or `X-shaped' dispersion around wave vector 
${\bf q}=(\pi,\pi)$\cite{Fujita} (which may also be  `Y-shaped'\cite{HgBaspin}).
This is frequently interpreted\cite{Fujita} as a magnon-like collective mode 
above the neck of the hour-glass co-existing with particle-hole excitations
of the fermi gas of free carriers below the neck. The part above the neck
of the hourglass thus would correspond to the triplet 
mode in Figure \ref{fig12}.
The mean-field treatment of the bond-particle Hamiltonian cannot reproduce the 
particle-hole excitations below the neck, but terms like 
${\bf t}_n^{}\; \cdot {\bf v}_{(n,\nu'),(m,\nu)}$ 
in (\ref{ham2}), which describe the decay of triplet into a particle-hole pair 
and which have been ignored in the mean-field treatment, may well produce 
these features. The mean-field triplet dispersion also does not reproduce the
paramagnon-excitations with ${\bf q}$ close to the zone 
center\cite{Paramagnon}.
These paramagnons show a decreasing frequency as 
${\bf q}\rightarrow 0$\cite{Paramagnon} which differs strongly from the
calculated $\omega_{\nu,{\bm q}}$. On the other hand in the mean-field
approximation we have neglected anharmonic terms such as
${\bf t}_n^\dagger \cdot ({\bf t}_m^\dagger\times {\bf t}_m^{})$
in (\ref{ham1}). By virtue of such terms a magnon with momentum
${\bf q}$ close to the zone center may decay into two magnons with momenta
$(\pi,\pi)+{\bf q}_1$ and $(\pi,\pi)+{\bf q}-{\bf q}_1$ with small
${\bf q}_1$ (coupled to a triplet) and the true magnetic excitation
may be a superposition of such states. We defer this to separate 
study, however.
\section{Summary}
In summary, we have presented a theory of the lightly doped paramagnetic
Mott-insulator by formulating the t-J Hamiltonian in terms of bond particles 
for a given dimer covering of the plane, and averaging this over coverings. 
The major simplification for low doping thereby comes about because the 
majority of dimers are assumed to be in the singlet state, which we 
re-interpreted as the 
vacuum state of the dimer, so that a theory for a low-density system of 
hole-like fermions and tripet-like bosons resulted.
By virtue of the low density, relaxing the infinitely strong repulsion between 
these remaining particles may be a reasonable approximation. In fact,
a similar approach has given reasonable results for the Kondo lattice, at least 
in the parameter range where the density of fermions and bosons indeed was 
small\cite{Kondo0,Kondo1,mykondo}.\\
The results describe what might be expected for a doped Mott insulator after 
long-range antiferromagnetic order has collapsed: due to their strong
Coulomb repulsion the electrons are `jammed' so that the all-electron fermi 
surface has collapsed. Instead, the electrons form an inert background - the 
`singlet soup' - and the only `active fermions' are the doped holes.
These correspond to spin-$\frac{1}{2}$ fermions and the fractional volume of 
the fermi surface is $x/2$ rather than $(1\pm x)/2$. As is the case in a 
Mott-insulator, the jammed electrons retain only their spin degrees of freedom 
and the exchange coupling between these results in a bosonic spin-triplet mode
with minimum at ${\bf q}=(\pi,\pi)$. The fermi surface consists of hole 
pockets centered near $(\frac{\pi}{2},\frac{\pi}{2})$ and symmetry equivalent 
points.
By and large, this description is consistent with a large body of experimental
results for the pseudogap phase of underdoped cuprates,
as discussed in the introduction. It should be stressed
that in order to obtain these results, use of the bond particle theory
is of advantage. Namely in the bond particle theory we
have hole-like spin-$\frac{1}{2}$ fermions (the $f_{m,\pm,\sigma}^\dagger$-fermions)
and triplet-like spin excitations (the ${\bf t}_m^\dagger$ bosons) as the
elementary excitations from the very outset. This can be contrated for example 
with the slave-boson/fermion representation of the t-J model where one
replaces $\hat{c}_{i,\sigma}^\dagger \rightarrow f_{i,\sigma}^\dagger b_i$
whereby the $b$-particle represents the empty site\cite{leeb}. In order to
model fermions which correspond to the doped holes one would have to
to assign fermi statistics to the $b$-particle, but then one would find 
spinless holes. On the other hand,
the proportionality $S=aT\chi$\cite{Loram} suggests
that the carriers in underdoped cuprates are spin-$\frac{1}{2}$ fermions.\\
Whereas the overall scenario predicted by the bond-particle theory
is consistent with experiment, a more detailed comparison shows
clear deficiencies.
Compared to experiment, the pocket is shifted towards $(\pi,\pi)$ and the 
width of the quasiparticle band is too large. The ${\bf k}$-dependence of 
the spectral weight at the fermi surface is too weak so that the spectral 
weight of the part of the pocket facing $(\pi,\pi)$ is too large to actually 
reproduce the fermi arcs. Moreover, the band width of the triplet bosons is
too small by a factor of $\approx 2$. On the other hand, this is only the 
mean-field result, where moreover the infinitely strong
repulsion between bond-particles has been simply neglected.
Taking this into account as well as the coupling between holes and 
triplets may result in modifications. This can be seen from the reasonably well 
understood problem of hole motion
in an antiferromagnet\cite{Bulaevskii,Trugman,Becker,MartinezHorsch}
where it is known that the hole is heavily dressed by spin fluctuations.
For spin-ladders the coupling between holes and triplets has
already been carried out\cite{Sushkovladder,JureckaBrenig} in the framework
of bond particle theory and given convincing results. Despite its obvious 
deficiencies the bond-particle formalism therefore may provide a reasonable 
starting point for more rigorous treatments of underdoped cuprates.

\end{document}